\newcommand{\xc}[1]{\textcolor{blue}{#1}}
\documentclass[conference]{IEEEtran}
\IEEEoverridecommandlockouts
\usepackage[font=footnotesize]{caption}  
\usepackage{graphicx} 
\usepackage{cite}
\usepackage{comment}
\usepackage{amsmath,amssymb,amsfonts}
\usepackage{booktabs}
\usepackage{subcaption}
\usepackage{wrapfig}
\usepackage{float}
\usepackage{parcolumns}
\usepackage{array}
\usepackage{multicol}
\usepackage{algorithm}
\usepackage{algpseudocode}
\usepackage{algorithmicx}
\usepackage{hyperref}
\usepackage{subcaption} 
\usepackage{booktabs}
\usepackage{mathtools}
\usepackage{tabularx}
\usepackage{xcolor}
\usepackage{bm}
\usepackage[normalem]{ulem}

\title{Fast-Response Variable-Frequency Series-Capacitor Buck VRM Through Integrated Control Approaches}
\author{%
  \IEEEauthorblockN{Guanyu Qian\thanks{This work was supported in part by the Faculty Startup Fund provided by the UCLA Samueli School of Engineering and Applied Science.}}
  \IEEEauthorblockA{%
    Electrical and Computer Engineering\\
    UCLA\\
    Los Angeles, USA\\
    gyqian@ucla.edu
  }
  \vspace{-25pt}
\and
  \IEEEauthorblockN{Haoxian Yan}
  \IEEEauthorblockA{%
    Electrical and Computer Engineering\\
    UCLA\\
    Los Angeles, USA\\
    leoyan@ucla.edu
  }
  \vspace{-25pt}

\and
  \IEEEauthorblockN{Xiaofan Cui}
  \IEEEauthorblockA{%
    Electrical and Computer Engineering\\
    UCLA\\
    Los Angeles, USA\\
    cuixf@seas.ucla.edu
  }
  \vspace{-25pt}
}

\begin{document}

\maketitle
\begin{abstract}
Fast-response voltage regulation is essential for data-center Voltage Regulation Modules (VRMs) powering Artificial Intelligence (AI) workloads, which exhibit both small-amplitude fluctuations and abrupt full-load steps.  This paper introduces a control scheme that integrates a linear controller and a nonlinear controller for variable-frequency Series-Capacitor Buck (SCB) converters.  First, an accurate small-signal model is derived via a Switching-Synchronized Sampled State-Space (5S) framework, yielding discrete-time transfer functions and root-locus insights for direct digital design.  A critical concern for SCB converters is series-capacitor oscillation during heavy load steps if the strict switching sequence is not maintained.  To accelerate large-signal transients, a time-optimal control strategy based on Pontryagin’s Maximum Principle (PMP) relaxes the switching constraints to compute time-optimal switching sequences. A transition logic is then proposed to integrate the high-bandwidth small-signal controller and the large-signal controller. Simulations demonstrate a fast output voltage recovery under a heavy load-step-up, over ten times faster than a linear-controller-only design. Preliminary hardware tests indicate a stable rejection to heavy load disturbances with zero steady-state error.
\\
\end{abstract}

\vspace{-5pt}
\vspace{-5pt}
\section{Introduction}
Approximately 4.4\% of U.S. electricity was consumed by microelectronics in data centers by 2023 \cite{shehabi_united_2016}, and this consumption is expected to continue increasing to meet the growing demand for Artificial Intelligence (AI) training and development. To regulate every watt that powers the Central Processing Units (CPUs) and Graphics Processing Units (GPUs), Voltage Regulation Modules (VRMs) are required to provide low output voltage and high current while maintaining a fast response speed to dynamic workloads\cite{li_ai_2025}. Consequently, VRM design for data centers must prioritize both efficiency and rapid transient response to highly variable load conditions.

The low voltage requirements make buck converter topology a reasonable choice, while their high current capability can be enhanced by paralleling and interleaving multiple phases. To balance the current between phases, a promising multiphase converter variant has been proposed by integrating a switched-capacitor voltage divider in each buck phase, known as the Series-Capacitor Buck Converter (SCB) topology \cite{shenoy_comparison_2015, Simon-muela_practical_2008,majumder_event-based_nodate}. Meanwhile, the modeling of the SCB suitable for fixed-frequency control approaches has been proposed in the \(s\)-domain, and the transient speed is fundamentally limited by the control bandwidth, which is approximately ten times lower than the switching frequency \cite{kirshenboim_closed-loop_2019}.

For a faster and more flexible control technique, the variable-frequency control scheme that utilizes Current-Mode Constant-On-Time (CM-COT) operation with digital implementation is especially appealing because it not only offers fast response and programmability but also aligns well with the recent resurgence of digitally controlled Power Management Integrated Circuits (PMICs)\cite{cho_fully_2022}. While the digital control enables scalable deployment, the CM-COT scheme offers benefits such as single-cycle current settling, faster transient response, and improved efficiency under light-load conditions \cite{cui_fast-response_2023}. However, accurate modeling suitable for fast variable frequency control has been lacking, and averaging strategies for variable-frequency converters, such as describing-function-based methods, are more complicated to apply for controller design \cite{jian_li_new_2010}. 

\begin{figure}[t]
    \centering
    \includegraphics[width=\linewidth]{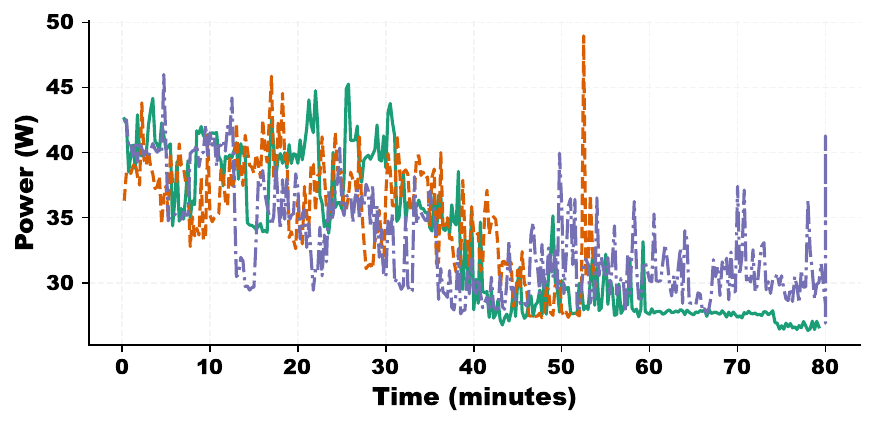}
    \caption{RTX 5080 GPU power consumption profiles during reinforcement‐learning training runs of the algorithm in \cite{mnih_human-level_2015}.}
    \label{fig:gpu_power}
    \vspace{-10pt}
\end{figure}

VRMs in data centers must accommodate both small-amplitude power perturbations and large-signal load transients, as well as dynamic voltage scaling (DVS) driven by abrupt AI training workload dynamics. Fig.~\ref{fig:gpu_power} illustrates a typical GPU power profile during reinforcement‐learning training, exhibiting both low‐magnitude fluctuations and load steps of nearly 100\%. Especially for the SCB converters, recent research\cite{naradhipa_partial_2023} highlights that undesired oscillations in the series capacitor voltage occur if the switching sequence during the transient events is not maintained, which imposes strict saturation bounds on the linear controller design using small-signal model and further degrades the transient performance of SCB VRMs compared to conventional buck VRMs, underscoring the need for effective control strategies for transient improvement.

This paper presents an integrated control approach for variable‐frequency SCB converters with the control algorithm implemented on a Field Programmable Gate Array (FPGA) platform.  An accurate small-signal model is derived within a Switching-Synchronized Sampled State-Space (5S) framework and validated through simulation, providing key insights for designing high-bandwidth small-signal controllers.  To address large-signal transients, an optimal control strategy based on Pontryagin’s Maximum Principle (PMP) is then developed, relaxing strict switching sequence constraints and permitting phase dephasing to accelerate the transient response. Building on these complementary approaches, strategies for seamlessly integrating the small-signal and PMP-based large-signal controllers are presented, followed by experimental validation of the small-signal controller performance and simulation results for the integrated scheme.
\section{Variable-Frequency SCB Converter using Event-Driven Digital Control}
\begin{figure}[t]
    \centering
    \includegraphics[width=0.96\linewidth]{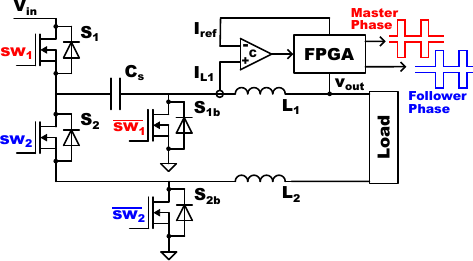}
    \caption{The schematic of the CM‐COT controlled two-phase series-capacitor buck converter with only master‐phase current feedback.}
    \label{fig:two-phase setup}
    \vspace{-10pt}
\end{figure}
\begin{figure*}[ht]
    \centering
    \includegraphics[width=0.96\linewidth]{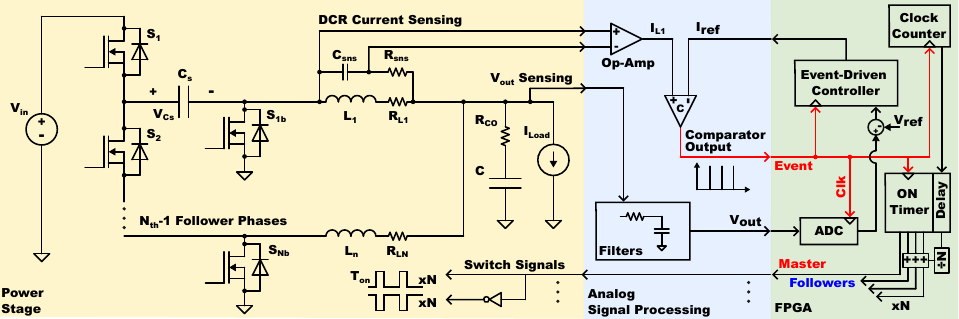}
    \caption{Schematic of the CM-COT multiphase SCB converter and control architecture.}
    \label{fig:setup}
    \vspace{-10pt}
\end{figure*}

\begin{algorithm}[h]
\caption{Switching-Synchronized Sampled PI Controller for Two-Phase Series-Capacitor Buck Converter}
\label{alg:scb_fpga}
\begin{algorithmic}[1]
  \Require $T_{\mathrm{on}}, T_{\mathrm{off,min}},k_p, k_i, V_{\mathrm{ref}} $\Comment{Constants}
  \Require \texttt{comparator\_out}, $v_{\mathrm{out}}$   \Comment{Inputs}
  \Ensure $I_{\mathrm{ref}},\;\texttt{sw}_1,\;\bar{\texttt{sw}}_1,\;\texttt{sw}_2,\;\bar{\texttt{sw}}_2$
  \Comment{Outputs}
    \State \textbf{Vars:} $e,\;\mathrm{int\_e},$ \Comment{Error Terms}
    \Statex \quad $t_d,\;\mathrm{period\_start},$ \Comment{Delay Calculation}
    \Statex \quad $v_{\mathrm{out\_sample},}$ \Comment{Voltage Sampler}
    \Statex \quad $\mathrm{period\_cnt},\;\mathrm{on\_cnt},\;\mathrm{off\_cnt}$ \Comment{Time Counters}

\State \textbf{Init:} 
  $(e,\;\mathrm{int\_e},\;\mathrm{period\_cnt},\;\mathrm{on\_cnt},\;\mathrm{off\_cnt})
    \!\gets\!0$
\Statex \quad 
  $(\texttt{sw},\;\bar{\texttt{sw}},\;\texttt{sw}_2,\;\bar{\texttt{sw}}_2)
    \!\gets\!(1,0,0,0)$
\Statex \quad 
  $\mathrm{period\_start}\!\gets\!\textbf{True}$, $v_{\mathrm{out\_sample}}\gets 0$ 
\Statex \quad
  \hrulefill\ \texttt{Controller Logic}\ \hrulefill
  \par\nobreak
  \vspace{2.5pt}
  \If{rising edge(\texttt{comparator\_out})}
    \State $v_{\mathrm{out\_sample}} \gets v_{\mathrm{out}}$ \Comment{Event-Driven Sample}
    \State $e \gets V_{\mathrm{ref}}-v_{\mathrm{out\_sample}}$, \quad $\mathrm{int\_e} \mathrel{+}= k_i\cdot e$
    \State $I_{\mathrm{ref}} \gets k_p \cdot e + \mathrm{int\_e}$
    \State $\mathrm{period\_start} \gets \lnot\,\mathrm{period\_start}$
    \State $t_d \gets \mathrm{period\_cnt} \,//\,2$ \Comment{Delay Holder}
    \State $(\mathrm{period\_cnt},\,\mathrm{on\_cnt},\,\mathrm{off\_cnt}) \gets 0$
  \EndIf
  \Statex \quad
  \hrulefill\ \texttt{Timer and Delay Manager}\ \hrulefill
  \par\nobreak
  \vspace{2.5pt}
  \State $\mathrm{period\_cnt} \mathrel{+}= \Delta t$ \Comment{Switching Period Tracker}
  \If{$\mathrm{on\_cnt}<T_{\mathrm{on}}$} 
    \State $\texttt{sw}_1\gets1,\;\mathrm{on\_cnt}\mathrel{+}=\Delta t$\Comment{Constant-On Timer}
  \ElsIf{$\mathrm{off\_cnt}<T_{\mathrm{off,min}}$} 
    \State $\texttt{sw}_1\gets0,\;\mathrm{off\_cnt}\mathrel{+}=\Delta t$ \Comment{Minimum-Off Timer}
  \Else
    \State $(\mathrm{on\_cnt},\,\mathrm{off\_cnt}) \gets 0$
  \EndIf
  
  \State $\bar{\texttt{sw}}_1 \gets \lnot\,\texttt{sw}_1$
  \State ${\texttt{sw}}_2 \gets \texttt{sw}_1$ delayed by $t_d$, \quad ${\bar{\texttt{sw}}}_2 \gets \bar{\texttt{sw}}_1$ delayed by $t_d$
\end{algorithmic}
\end{algorithm}
In the following sections, the 5S framework is applied to analyze a two-phase SCB converter with the CM-COT modulation method (COT-CM SCB) to achieve a high-performance digital control, and the top-level schematic is shown in Fig.~\ref{fig:two-phase setup}.  The sampling action in this framework is event-driven and in synchronization with the variable switching frequency \cite{cui_fast-response_2023}. 
Therefore, we need to examine the signals that require sampling and identify the corresponding sampling events. The output voltage is one such signal and must be fed back to the controller. Moreover,  for the current feedback and sensing, the SCB's automatic current-balancing mechanism ensures that, in the steady state, all follower-phase valley currents match the reference current, \(I_{\text{ref}}\), as long as the master-phase's valley current, \(i_{L1}\), is regulated to \(I_{\text{ref}}\). Consequently, only the master-phase current needs to be measured and fed back, which not only reduces the number of required current sensors but also simplifies the controller design. 
The comparator output thus qualifies as the sampling event. When the measured current \(I_{L1}\) falls below \(I_{\text{ref}}\), the comparator output transitions from low to high, simultaneously triggering voltage sampling and initiating the next on-time cycle. 
Fig.~\ref{fig:setup} presents the complete schematic of the SCB, incorporating a switching-synchronized sampled state space control architecture at the circuit level that can be potentially scaled up for multiphase operation, where the processor load is modeled as a current sink at the output~\cite{naradhipa_partial_2023}. 

With the sampling events and feedback signals defined, the switching-synchronized controller logic and phase‐delay management can now be established. During each switching cycle, the controller's output is updated once using the sampled output voltage information to generate the new reference current command. Then, phase management in the SCB is implemented as a simple delay line: the comparator trigger event initiates a delay counter in each follower phase. For the \(l_\text{th}\) follower phase, the relative delay in \(n_\text{th}\) cycle, is defined by \(t_{d,l}[n] = T_{\text{sw}}[n-1]/N \) where \( T_{\text{sw}}[n-1]\) is the master‐phase switching period in the cycle \((n-1)\)th cycle and \(N\) is the total number of phases. Notably, the complete FPGA implementation, including the digital Proportional Integral (PI) controller for two‐phase SCB operation, requires only twenty lines of code (see Algorithm~\ref{alg:scb_fpga}).
\section{Small-Signal Modeling and Linear Controller Design of SCB Converters}
Building on the preceding control architecture and implementation, a small-signal model for the SCB converter can be derived within the 5S framework, which facilitates discrete-domain transfer functions that aid in designing high-bandwidth controllers in the variable-frequency scheme.

\subsection{Modeling}
Within this framework, the relationship between the output voltage \(v[n]\) and the master‐phase inductor current \(i_{L1}[n]\) is governed by a nonlinear difference equation derived from the time‐domain waveform. To simplify notation, all phase inductances are set to be identical (\(L_1 = L_2 = \cdots = L_N = L\)). Two key assumptions are then made for the SCB converter:
\begin{enumerate}
  \item The series capacitor is sufficiently large to be modeled as an ideal voltage source in the steady state with negligible ripple and zero equivalent series resistance (the same approximation also applies to the output capacitor).
  \item All inductor currents follow fixed linear‐ramp ramps within each subinterval of the switching cycle, characterized by slopes \(m_1^l\) or \(m_2^l\), where \(l=0\) denotes the master phase and \(l=1,\dots,N-1\) the follower phases.  Under the ideal‐voltage‐source approximation, these slopes are identical across all phases:
  \[
    m_1^l = m_1 = \frac{V_{\mathrm{in}}/2 - V_{\mathrm{out}}}{L},
    \quad
    m_2^l = m_2 = \frac{V_{\mathrm{out}}}{L},
    \quad\forall\,l.
  \]
\end{enumerate}

\begin{figure}[ht]
    \centering
    \includegraphics[width=0.96\linewidth]{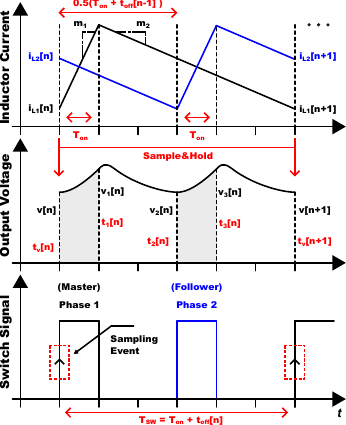}
    \caption{Time-domain waveforms of the master-phase inductor current, output voltage, and switch signal, with the sampling event indicated.}
    \label{fig:waveform}
    \vspace{-10pt}
\end{figure}

For clarity, a two-phase SCB is demonstrated with detailed procedures in this paper. Fig.~\ref{fig:waveform} presents the annotated time-domain waveforms over one switching period defined by the sampling event at time instants \(t_v[n]\) and \(t_v[n+1]\). This period is divided into four subintervals:
\[
\begin{aligned}
\text{Interval 1:}&\ [\,t_v[n],\,t_1[n]\,], \quad &\text{Interval 2:}&\ [\,t_1[n],\,t_2[n]\,],\\
\text{Interval 3:}&\ [\,t_2[n],\,t_3[n]\,], \quad &\text{Interval 4:}&\ [\,t_3[n],\,t_v[n+1]\,].
\end{aligned}
\]

Intervals 1 and 3 each have fixed duration \(T_{\mathrm{on}}\); the duration of Interval 2, \(t_2[n]-t_1[n]\), is set by the previous cycle, and Interval 4, \(t_v[n+1]-t_3[n]\), is defined by the remaining off‐time within the current switching period.
Under the piecewise‐linear assumptions, the discrete‐time inductor currents for the master phase, \(i_{L1}[n]\), and the follower phase, \(i_{L2}[n]\), are updated at each sampling instant according to the following sequence:
\begin{align}
\label{eq:iL1}
i_{L1}[n+1] &= i_{L1}[n]
  + m_1\,T_{\mathrm{on}}
  - m_2\,t_{\mathrm{off}}[n],\\[6pt]
\label{eq:iL2}
i_{L2}[n+1] &= i_{L2}[n]
  + m_1\bigl(t_1[n] - t_v[n]\bigr)\notag\\
&\quad\;
  -\,m_2\bigl(t_2[n] - t_v[n] + t_v[n+1] - t_3[n]\bigr).
\end{align}
where \(T_{\mathrm{on}}\) is held constant by the COT scheme, and \(t_{\mathrm{off}}[n]\) is set by the current sample \(i_{L1}[n]\) and future sample \(i_{L1}[n+1]\).  

With the inductor‐current update laws established and under assumption (2), the output capacitor current also follows a piecewise‐constant ramp, given by the net inductor currents minus the load current.  In particular, the capacitor current at the start of cycle \(n\) is
\begin{equation}
    i_c[n] = i_{L1}[n] + i_{L2}[n] - I_{\mathrm{load}},
\end{equation}
and the corresponding capacitor‐voltage evolution over the four subintervals is given by:
\begingroup
  \setlength{\jot}{8pt} 
  \begin{align}
    v_1[n] &= v[n]
             + \frac{T_{\mathrm{on}}}{C}\,i_c[n]
             + \frac{T_{\mathrm{on}}^{2}}{2C}\,(m_1 - m_2),\\
    v_2[n] &= v_1[n]
             + \frac{t_2[n] - t_1[n]}{C}\bigl(i_c[n] + T_{\mathrm{on}}(m_1 - m_2)\bigr)\notag\\
           &\quad - \frac{(t_2[n] - t_1[n])^{2}}{2C}\,(2m_2),\\
    v_3[n] &= v_2[n]
             + \frac{T_{\mathrm{on}}}{C}\Bigl(i_c[n] + T_{\mathrm{on}}(m_1 - m_2)\notag\\
           &\quad - 2\,m_2\,(t_2[n] - t_1[n])\Bigr)
             + \frac{T_{\mathrm{on}}^{2}}{2C}\,(m_1 - m_2),\\
    v[n+1] &= v_3[n]
             + \frac{t_v[n+1] - t_3[n]}{C}\bigl(i_c[n] \notag\\
           &\quad + 2\,T_{\mathrm{on}}(m_1 - m_2)
                   - 2\,m_2\,(t_2[n] - t_1[n])\bigr)\notag\\
           &\quad - \frac{(t_v[n+1] - t_3[n])^{2}}{2C}\,(2m_2).
  \end{align}
\endgroup

By applying small-signal perturbation and linearization, the relation between \(v[n]\) and \(i_{L1}[n]\) can be expressed as
\begin{align}
\tilde v[n+1] &= \alpha\,\tilde i_{L1}[n]
             + \beta\,\tilde i_{L2}[n]
             + \gamma\,\tilde i_{L1}[n-1] \nonumber\\
             &\quad
             + \eta\,\tilde i_{L1}[n+1]
             + \theta\,\tilde i_{\mathrm{Load}}[n]
             + \tilde v[n],
\end{align}
with the following parameters:
\begin{align}
\alpha &= 2K(2M+1),      & \beta  &= 2K(M+1),      \notag\\
\gamma &= -K(M+1),       & \eta   &= K(1 - M),     \notag\\
\theta &= -2K(M+1),      & K      &= T_{\mathrm{on}}/{2\,C},\quad
M      = 2\,V_{\mathrm{out}}/{V_{\mathrm{in}}}.\notag
\end{align}

The dependence of \(\tilde v[n+1]\) on \(\tilde i_{L1}[n+1]\) appears because the off‐time \(t_{\mathrm{off}}[n]\) is determined by the future master‐phase current sample \(i_{L1}[n+1]\).  Likewise, the historical term in \(\tilde i_{L1}[n-1]\) enters via the follower‐phase current \(\tilde i_{L2}[n]\), which is delayed by one cycle relative to the master phase.  
It is worth noting that, under the CM-COT control scheme, the master-phase inductor current \(i_{L1}[n]\) is precisely regulated to match the reference current \(I_{\text{ref}}[n]\) at each sampling event. Moreover, the reference current \(I_{\text{ref}}[n]\)—as the output of the controller is equivalent to the system input \(u[n]\). Therefore \(u[n]\) is directly related to the next current value:
\begin{equation}
    u[n] = I_{\mathrm{ref}}[n] = i_{L1}[n+1].
\end{equation}

Furthermore, by defining a new state variable: \(i_{Ld}[n] = i_{L1}[n-1]\), the small-signal system dynamics can be written in a standard state-space form:  
\begin{equation}
\begin{aligned}
\underbrace{%
\begin{bmatrix}
\tilde v[n+1]\\
\tilde i_{L1}[n+1]\\
\tilde i_{L2}[n+1]\\
\tilde i_{Ld}[n+1]
\end{bmatrix}}_{\mathbf{x}[n+1]}
&=
\underbrace{%
\begin{bmatrix}
1 & \alpha & \beta & \gamma \\
0 & 0 & 0 & 0\\
0 & -1 & 1 & 0\\
0 & 1 & 0 & 0
\end{bmatrix}}_{A}
\underbrace{%
\begin{bmatrix}
\tilde v[n]\\
\tilde i_{L1}[n]\\
\tilde i_{L2}[n]\\
\tilde i_{Ld}[n]
\end{bmatrix}}_{\mathbf{x}[n]}\\[6pt]
\quad
&+ \underbrace{%
\begin{bmatrix}
\eta\\
1\\
1\\
0
\end{bmatrix}}_{B_u}\,\tilde u[n]
+ \underbrace{%
\begin{bmatrix}
\theta\\
0\\
0\\
0
\end{bmatrix}}_{B_d}\,\tilde i_{\mathrm{Load}}[n]
\end{aligned}
\end{equation}
\begin{equation}
\underbrace{%
\tilde v[n]
}_{\mathbf{y}[n]}
=
\underbrace{%
\begin{bmatrix}
1 & 0 & 0 & 0
\end{bmatrix}}_{C}
\mathbf{x}[n],
\end{equation}
the \(z\)-domain transfer function from \(\tilde{u}(z)\) to \(\tilde{v}(z)\) can be obtained through \(C(zI-A)^{-1}B_u\), yielding:
\begin{equation}
    \frac{\tilde v(z)}{\tilde {u}(z)} = \frac{T_{\mathrm{on}}}{2C\cdot M}\frac{(1-2M)z^2+(4+2M)z-1}{z^2(z-1)}.
    \label{eq:vtoitf}
\end{equation}
\begin{figure}[t]
    \centering
    \includegraphics[width=\linewidth]{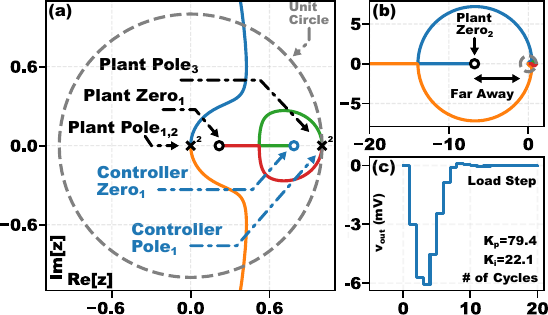}
    \caption{(a) Closed‐loop root locus based on Eqs.~\eqref{eq:vtoitf} and~\eqref{eq:pi}. (b) One of the plant zeros is located outside the unit circle. (c) Output voltage recovery for a 1 A load step with the designed controller.}
    \label{fig:root_locus}
    \vspace{-10pt}
\end{figure}
\begin{figure*}[t]
    \centering
    \includegraphics[width=0.96\linewidth]{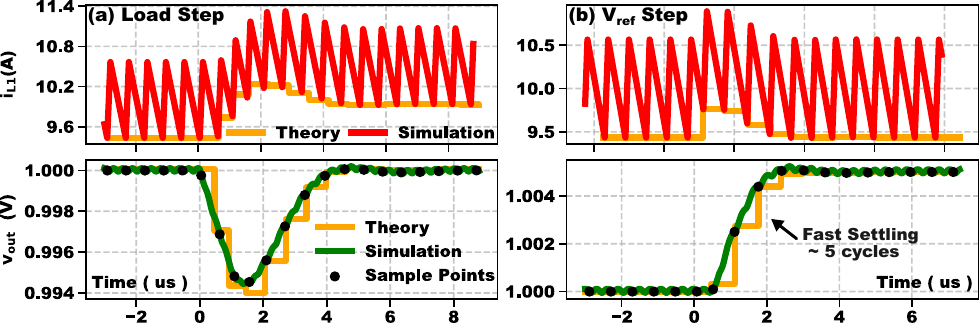}
    \caption{(a) Master‐phase current and output voltage responses for a 1 A load‐step. (b) Current and voltage responses for a 5 mV reference‐step. Sampled points and theoretical current trajectories are annotated.}
    \label{fig:simu_theory}
    \vspace{-10pt}
\end{figure*}
The small-signal transfer function exhibits no explicit dependence on the inductance \(L\). This arises from approximating each inductor current as a time-invariant linear ramp. The accuracy trade-offs associated with this approximation are well-discussed in \cite{cui_fast-response_2023}. For the two-phase SCB converter with parameters listed in Table~II, treating inductor current as a fixed linear ramp preserves modeling accuracy while keeping computational and modeling complexity at a moderate level.

Similarly, by selecting  \(C   = [0 \quad 0 \quad 1 \quad 0],\) the small‐signal transfer function from \(\tilde u(z)\) to \(\tilde i_{L2}(z)\) becomes  
\begin{equation}
\frac{\tilde i_{L2}(z)}{\tilde u(z)} = \frac{1}{z},
\label{eq:currentdelay}
\end{equation}
which is exactly a one-cycle delay. In addition, the Eq~\eqref{eq:vtoitf} can expressed in zero–pole–gain (ZPK) form as:
\begin{equation}
    \frac{\tilde v(z)}{\tilde i_{L1}(z)}
= K \,\frac{(z - z_1)(z - z_2)}{z^2 (z - 1)}
\end{equation}
\begin{equation}
    z_{1,2}= \frac{M + 2 \pm \sqrt{M^2 + 2M + 5}}{2M - 1},
    \label{eq:zero}
\end{equation}
providing clear insights into the converter’s dynamics:
\begin{enumerate}
  \item One pole at \(z\) = 0 arises from the causality nature of the discrete-control; the current-cycle control input aims to impact the next-cycle inductor current, same as the buck converter\cite{cui_fast-response_2023}. The extra pole at \(z\) = 0 originates from a one-cycle delay between the follower and master phases, as seen in Eq.~\eqref{eq:currentdelay}.
  \item The pole at \(z=1\) corresponds to the integrating behavior of the output capacitor; in practice, this pole shifts slightly inside the unit circle due to parasitic resistance in the circuit.
  \item The zeros \(z_{1,2}\) are all real numbers and determined solely by the DC conversion ratio \(M\).
\end{enumerate}

To highlight the final point, consider a typical VRM buck converter with a step-down ratio larger than 3; it can be proven that the zeros must have different signs, with a zero location \( \in (0.21, 0.22) \) and another zero location \(\in (-14.21, -4.24)\). Both zeros are away from the unit circle, therefore having a minimum impact on the dominant dynamics. 

The left‐half‐\(z\)-plane (LHZP) zero originates from the discrete‐time feedforward of an inductor‐current perturbation into the capacitor voltage within a single switching period \cite{cui_fast-response_2023}.  Although a capacitor voltage cannot change instantaneously in continuous time when the inductor current is perturbed, in the discrete-time framework, voltage and current variations co-occur during the same cycle.

The right-half-\(z\)-plane (RHZP) zero indicates the non-minimum phase behavior inherent to CM-COT operation: as the inductor current increases, the off-time reduces, shortening the switching cycle and delivering less charge to the output capacitor.
\subsection{Controller Design}
Classical controller design methods may then be applied to design a high‐speed, digital PI controller in the form of 
\begin{equation}
C(z) \;=\; k \,\frac{z - z_k}{\,z - p_k\,}.
\label{eq:pi}
\end{equation}

Using the parameters in Table~II with \(I_{\mathrm{load}} = 20\)\, A, Fig.~\ref{fig:root_locus}(a–c) shows the root‐locus plots for the load‐step design with a digital PI controller; the reference‐step design follows similarly.  The open‐loop transfer function exhibits three poles—two at \(z=0\) and one at \(z=1\).  We focus on the dominant closed‐loop poles that originate from \(z=1\).  The first design parameter is the controller zero \(z_k\), which is placed near \(z=1\) to compensate for the slow output‐capacitor pole. 
Effective compensation reduces the dominant pole pair to a single locus moving from the controller pole at \(z=1\) toward the plant zero \(z_1\in(0.21,0.22)\).  The second design parameter is the controller gain \(k\), whose upper bound is set so that the two poles originating from \(z=0\) remain at sufficiently high frequency (\(\lVert z\rVert<0.1\)), and  Fig.~\ref{fig:root_locus}(d–e) presents the closed‐loop responses for the selected gain.  

Simulations of a 5 mV reference step and a 1 A load step in PLECS confirm that the theoretical time‐domain responses closely match the simulated results (Fig.~\ref{fig:simu_theory}(a–b)), with the optimized controller achieving settling in approximately five switching cycles.

\subsection{Saturation Boundary and Limitations}

In practical applications—ranging from AI training to inference—the processor load current can fluctuate by nearly 100\% \cite{li_ai_2025,10.1007/978-3-031-23220-6_8}, which can break the small-signal model’s assumptions (Section~III-A), leading to large voltage deviations and hitting the minimum-off-time limit.  Moreover, a non-overlap turn-on between \(S_1\) and \(S_2\)  must be enforced \cite{naradhipa_partial_2023} to prevent the series‐capacitor from overcharging.

Enforcing such non-overlap turn-on law in a purely linear design requires a minimum off‐time  
\(T_{\mathrm{off,min}} =  T_{\mathrm{sw}}/2,\)
which eliminates overcharging but limits the maximum transient slew rate. For instance, during a 10 A load‐step up, each cycle's off-time saturates at \(T_{\mathrm{off,min}}\) (Fig.~\ref{fig:compare_all}(b)). Under strict interleaving, the master phase must first charge up the series capacitor before any energy can flow through the follower phase to the output, thereby slowing the current ramp-up.  These saturation limits highlight the need for advanced control strategies to achieve faster large‐signal performance in SCB converters.  

\section{Time-Optimal Trajectory-Based Control}
\begin{figure*}[t]
    \centering
    \includegraphics[width=\linewidth]{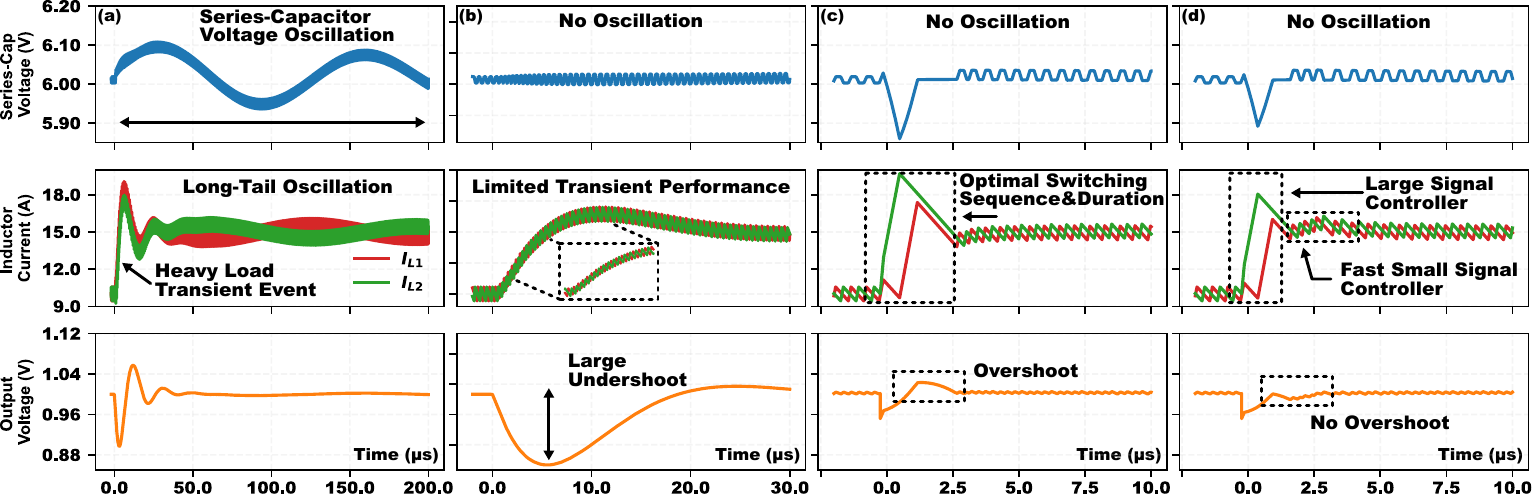}
    \caption{Transient comparison: (a) capacitor oscillation when the non‐overlap constraint is violated (both top switches on simultaneously); (b) linear‐controller response enforcing strict sequence via \(T_{\mathrm{off,min}}\); (c) large‐signal controller with time‐optimal sequence with precise mode transitions; (d) integrated large‐signal and small‐signal controller response.}
    \label{fig:compare_all}
    \vspace{-10pt} 
\end{figure*}

While treating the series capacitor as an ideal voltage source is valid in the steady state, it does not hold well for the heavy load transients. Specifically, if switches \(S_1\) and \(S_2\) conduct simultaneously, the input source can overcharge the series capacitor, causing its voltage to deviate and requiring a prolonged recovery as shown in Fig.~\ref{fig:compare_all}(a). This also serves as one of the reasons why the non-overlap law has to be maintained in \cite{naradhipa_partial_2023}. Accordingly, a large‐signal controller is introduced to address this transient limitation. By relaxing the non-overlap constraint and permitting both top switches to turn on, the effective energy-transfer path is shortened, allowing follower inductor currents to accelerate the transient response. However, the concurrent switch turn-on event not only charges up the capacitor quickly but also alters each phase’s current‐ramp slope. Transient switching sequences and durations are hard to design intuitively. To resolve this, we formulate the SCB as a switched-affine system and employ optimal control theory to compute the switching sequence and timing that minimize the load-step response time \cite{noauthor_time_1999}.
\subsection{Modes and Formulation}

Let \(x_0\) denote the initial state (e.g., at the onset of a load step) and \(x_f\) denote the desired final state, which is accompanied by a finite tolerance \(\epsilon\).  The time‐optimal control problem can then be stated as:
\begin{align}
\underset{T_1,\dots,T_{2^N}}{\mathrm{minimize}}\quad 
  & J = T_f = \sum_{i=1}^{2^N} T_i,\notag \\[6pt]
\text{subject to}\quad
  & T_i \ge 0,\quad i = 1,\dots,2^N,\quad  x_0,\,x_f, \, \epsilon\in \mathbb{R}^{2^N},\notag \\[4pt]
  & x(0) = x_0, \quad  x(T_f) = x_f + \epsilon.
  \label{eq:obj}
\end{align}
where \(N\) refers to the number of phases; Table~I shows the mode definition for a two-phase SCB converter.  
\begin{table}[h]
  \centering
  {\fontsize{8pt}{10pt}\selectfont
   \renewcommand{\arraystretch}{1}%
   \caption{Switching modes for a two‐phase SCB converter.}
   \label{tab:modes}
   \begin{tabularx}{0.48\textwidth}{@{}l 
       >{\centering\arraybackslash}X 
       >{\centering\arraybackslash}p{1.5cm}@{}}
     \toprule
     \textbf{Mode}
     & \(\bm{(S_{1},\,S_{1b},\,S_{2},\,S_{2b})}\)
     & \textbf{Duration} \\
     \midrule
     Mode 1 & \((1,0,1,0)\) & \(T_1\)  \\
     Mode 2 & \((1,0,0,1)\) & \(T_2\)  \\
     Mode 3 & \((0,1,1,0)\) & \(T_3\)  \\
     Mode 4 & \((0,1,0,1)\) & \(T_4\)  \\
     \bottomrule
   \end{tabularx}
  }
\end{table}

The converter's state dynamics can then be expressed with the following variables in the state‐space format:
\begin{align}
    x(t) &= [i_{L1}(t) \quad i_{L2}(t)\quad v_{cs}(t)\quad v_{out}(t)]^T,\\ 
    \dot x(t) &= A^{(k)}\,x(t) + B^{(k)}, \quad k \in \{1,2,3,4\},
\end{align}
where \(A^{(1\text{–}4)}\) and \(B^{(1\text{–}4)}\) correspond to the four switching modes.  Notably, the equivalent series resistance \(R_{\mathrm{co}}\) of the output capacitor is intentionally introduced as a transient‐mode detector, and the MOSFET on‐resistance \(R_{\mathrm{ds}}\) is retained not only for its impact on overall converter efficiency but also for its role as a damping resistor on the series capacitor, thereby influencing the series capacitor’s state dynamic. Admittedly, since only \(v_{out}\) and \(i_{L1}\) are directly measured, a full‐order state observer is required to estimate \(i_{L2}(0)\) and \(v_{cs}(0)\) from the available measurements and the known state‐space dynamics.  

{\small
\renewcommand{\arraystretch}{-0.3}    
\setlength{\arraycolsep}{6pt}         

\newcommand{\fixedAmat}[1]{%
  \resizebox{0.42\textwidth}{!}{$\displaystyle#1$}%
}

\[
\begin{aligned}
A^{(1)} &=
\fixedAmat{
  \begin{bmatrix}
   R_{\mathrm{ds}}/L & 0 & -1/L & -1/L\\[8pt]
   0 & -R_{\mathrm{ds}}/L & 0 & -1/L\\[8pt]
   1/C_{s} & 0 & 0 & 0\\[8pt]
   (1/C) - (R_{co}R_{\mathrm{ds}})/L
     & (1/C) - (R_{co}R_{\mathrm{ds}})/L
     & -R_{co}/L
     & -2R_{co}/L
  \end{bmatrix}
}, 
\\[4pt]
B^{(1)} &=
\begin{bmatrix}
 V_{\mathrm{in}}/L
 & V_{\mathrm{in}}/L
 & 0
 & -I_{\mathrm{load}}/C + 2R_{co}V_{\mathrm{in}}/L
\end{bmatrix}^T;
\\[14pt]
A^{(2)} &=
\fixedAmat{
  \begin{bmatrix}
   -R_{\mathrm{ds}}/L & 0 & -1/L & -1/L\\[8pt]
   0 & -R_{\mathrm{ds}}/L & 0 & -1/L\\[8pt]
   1/C_{s} & 0 & 0 & 0\\[8pt]
   (1/C) - (R_{co}R_{\mathrm{ds}})/L
     & (1/C) - (R_{co}R_{\mathrm{ds}})/L
     & -R_{co}/L
     & -2R_{co}/L
  \end{bmatrix}
}, 
\\[4pt]
B^{(2)} &=
\begin{bmatrix}
 V_{\mathrm{in}}/L
 & 0
 & 0
 & -I_{\mathrm{load}}/C + R_{co}V_{\mathrm{in}}/L
\end{bmatrix}^T;
\\[14pt]
A^{(3)} &=
\fixedAmat{
  \begin{bmatrix}
   -R_{\mathrm{ds}}/L & 0 & 0 & -1/L\\[8pt]
   0 & -R_{\mathrm{ds}}/L & 1/L & -1/L\\[8pt]
   0 & -1/C_{s} & 0 & 0\\[8pt]
   (1/C) - (R_{co}R_{\mathrm{ds}})/L
     & (1/C) - (R_{co}R_{\mathrm{ds}})/L
     & R_{co}/L
     & -2R_{co}/L
  \end{bmatrix}
}, 
\\[4pt]
B^{(3)} &=
\begin{bmatrix}
 0 & 0 & 0 & -I_{\mathrm{load}}/C
\end{bmatrix}^T;
\\[14pt]
A^{(4)} &=
\fixedAmat{
  \begin{bmatrix}
   -R_{\mathrm{ds}}/L & 0 & 0 & -1/L\\[8pt]
   0 & -R_{\mathrm{ds}}/L & 0 & -1/L\\[8pt]
   0 & 0 & 0 & 0\\[8pt]
   (1/C) - (R_{co}R_{\mathrm{ds}})/L
     & (1/C) - (R_{co}R_{\mathrm{ds}})/L
     & 0
     & -2R_{co}/L
  \end{bmatrix}
}, 
\\[4pt]
B^{(4)} &=
\begin{bmatrix}
 0 & 0 & 0 & -I_{\mathrm{load}}/C
\end{bmatrix}^T.
\end{aligned}
\]
}
\subsection{Hamiltonian and costate}
For each mode \(k\), the Hamiltonian system can be written as follows with a one-hot vector \( i_{k(t)} \in \{0,1\}^s\) and a mode indicator variable \(s\):
\begin{align}
\dot x &= \frac{\partial \mathcal{H}}{\partial \lambda}
       = \sum_{k=1}^{s}i_{k(t)}( A^{(k)}\,x + B^{(k)})
       \label{eq:H_start}
\\
\dot\lambda &= -\frac{\partial \mathcal{H}}{\partial x}
             = -\sum_{k=1}^{s}i_{k(t)}\,A^{(k)T}\,\lambda,
\end{align}
where \(\lambda\) is the co-state, and the Hamiltonian function can be expressed as:
\begin{equation}
    \mathcal{H}_k(x,\lambda,i) = \sum_{k=1}^{s}i_{k(t)} \lambda^T \bigl(A^{(k)} x + B^{(k)}\bigr), 
\end{equation}

The necessary condition for the \(x(t)\) with free \(T_f\) but fixed terminal state to have an optimal time trajectory to exist is that the costate function is absolutely continuous and never zero, while the following two conditions are satisfied \cite {noauthor_time_1999}
\begin{equation}
    \mathcal{H}_k(x,\lambda) =  \max_{\,k\in\{1,\dots,S\}}  \lambda^T \bigl(A^{(k)} x + B^{(k)}\bigr), 
\end{equation}
\begin{equation}
    \mathcal{H}_k(x,\lambda) \ge 0 \;\text{on}\; [0, T_f] 
    \label{eq:H_end}
\end{equation}

It's worth noting that the \(\lambda\) can be solved numerically through the shooting method and the bisection line search method, and the switching actions occur at:
\begin{equation}
    \lambda^{T} \bigl(A^{(k^*)} x + B^{(k^*)}\bigr)  \ge \lambda^T \bigl(A^{(k)} x + B^{(k)}\bigr), 
\end{equation}

\subsection{Numerical Results and Controller Integration}
The time‐optimal problem was solved numerically using the parameters in Table~II, including \(R_{\mathrm{co}} = 5\:\mathrm{m}\Omega\) and \(R_{\mathrm{ds}} = 2.2\:\mathrm{m}\Omega\). We first employed a nonlinear optimization solver on Eq.~\eqref{eq:obj}, leveraging the fact that linear mode dynamics allow consecutive intervals with the same \((A^{(k)},B^{(k)})\) to be merged—reducing the search to at most \(s!\) distinct mode and yielding fast runtimes. A PMP-based search, based on Eqs.~\eqref{eq:H_start}–\eqref{eq:H_end}, was then used to verify these results, with full agreement. The resulting optimal switching sequence and durations for a 10 A step up (20 A$\rightarrow $30 A) are  
\[
\text{Sequence: }[1,\,3,\,2,\,4], 
\quad
T_i = [101,\,589,\,629,\,1045]\:\mathrm{ns}.
\]

Fig.~\ref{fig:compare_all}(c) illustrates that the time-optimal controller achieves a settling time of approximately \(2.5\,\mu\text{s}\), in contrast to \(30\,\mu\text{s}\) for the pure small-signal design shown in Fig.~\ref{fig:compare_all}(b). In this design, the tolerance parameter \(\epsilon\) was initially set to a very small value to ensure precise mode transitions. In practice, \(\epsilon\) can be adjusted to enhance robustness while minimizing oscillatory behavior. A near-zero tolerance for the series-capacitor voltage helps prevent oscillatory charging, whereas a moderate tolerance for other state variables allows the small-signal controller to take over seamlessly, which is referred to as the integrated scheme. 
Fig.~\ref{fig:compare_all}(d) demonstrates that the integrated scheme not only eliminates overshoot but also achieves small-signal settling in roughly five switching cycles. This is accomplished while preserving the overall recovery time of \(2.5\,\mu\text{s}\), given that the final inductor current values in both phases are estimated with 10 percent uncertainty. In the load‐step‐up transient, the time-optimal controller will let both phase currents briefly increase at the maximum slew rate, a behavior typically acceptable given MOSFETs’ ability to tolerate high pulse currents (hundreds of amperes). If tighter current limits or protections are required, one can impose additional constraints on the dwell times, e.g., \(0 \le T_i \le T_{\max}\), to prevent excessive switch turn-on time.

\begin{figure}
    \centering
    \includegraphics[width=0.96\linewidth]{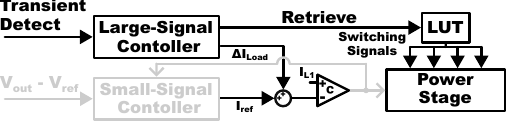}
    \caption{FPGA integration logic for large-signal and small-signal controllers. Upon detecting a heavy load via \(R_{\mathrm{co}}\), the FPGA calculates the step magnitude, freezes the small-signal loop, and adds this value to \(I_{\mathrm{ref}}\). It then retrieves the precomputed, time-optimal switching sequence from the LUT and sends it to the power stage.}

    \label{fig:fpga_integrate}
    \vspace{-10pt}
\end{figure}

For FPGA implementation, the optimal switching sequence and dwell durations are precomputed and stored in a Look‐Up Table (LUT). Fig.~\ref{fig:fpga_integrate} depicts the integration logic for a load‐step event. Upon transient detection, the small‐signal controller freezes its internal state and ceases updates. The large-signal controller then adds the estimated step magnitude to the reference current and holds the delay counter at its previous cycle value. Next, the FPGA retrieves the optimal sequence and timings from the LUT and sends them as switching signals to the power stage. Once the sequence completes, the comparator output is forced high to reset the sampling event, and normal controller operation resumes.  
\section{Experimental Results}

\begin{figure}
    \centering
    \includegraphics[width=0.96\linewidth]{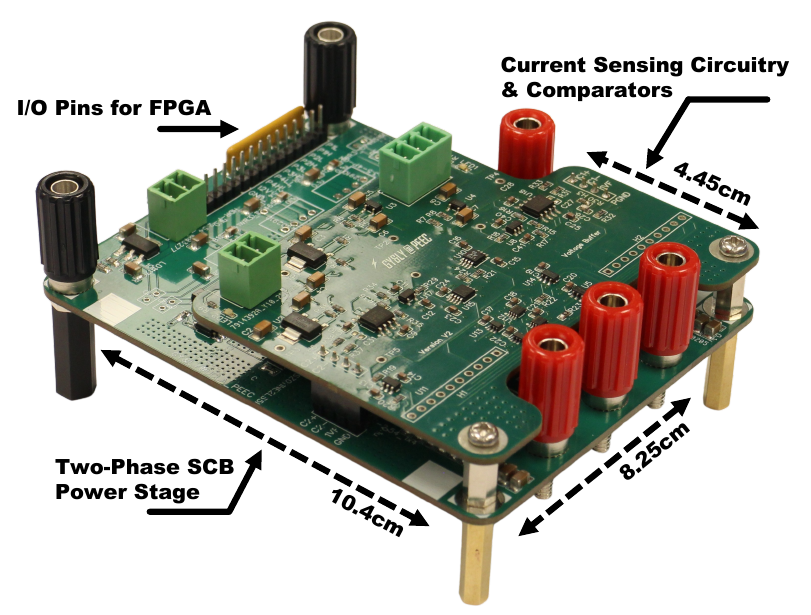}
    \caption{Circuit setup of a two-phase series-capacitor buck
        converter with current sensing and comparators.}
    \label{fig:circuit}
    \vspace{-15pt}
\end{figure}

\begin{figure}[t]
    \centering
     \includegraphics[width=0.96\linewidth]{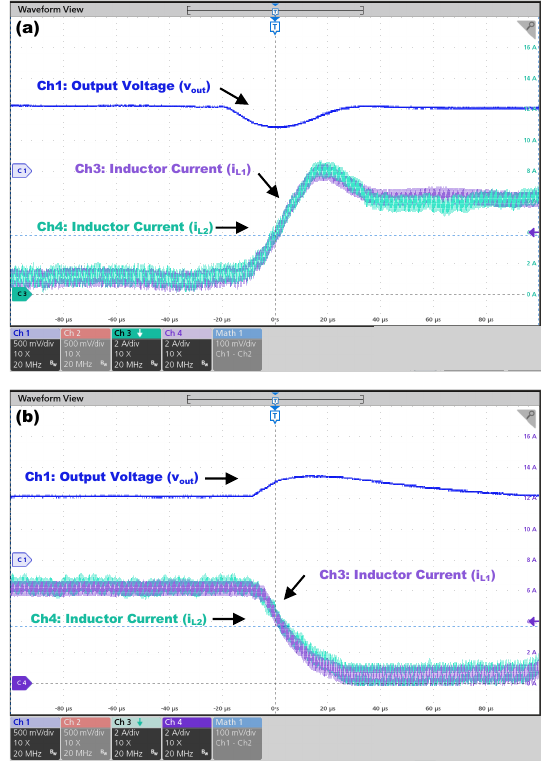}
    \caption{Experimental waveforms of output voltage and inductor currents during a 12 A load‐step (a) up and (b) down.}
    \label{fig:exp_step}
    \vspace{-15pt}
\end{figure}

A two-phase SCB prototype shown in Fig.~\ref{fig:circuit} was built and tested using the parameters listed in Table II. The experimental waveform—measured at 1.67 MHz with a 12 A load‐step up and down, shows a response time of approximately \(30\mu s\)  (Fig.~\ref{fig:exp_step}), underscoring the limitations of a purely linear‐controller design and providing experimental validation for Fig.~\ref{fig:compare_all}(b).
\begin{table}[h]
  \centering
  \caption{Design Parameters of the CM‐COT Converter}
  \label{tab:cotcm_buck_param}
  {\fontsize{8pt}{10pt}\selectfont                   
   \renewcommand{\arraystretch}{1}
   \begin{tabular*}{0.48\textwidth}{@{\extracolsep{\fill}} 
       l c l c @{}}
     \toprule
     \textbf{Parameter} & \textbf{Value} & \textbf{Parameter} & \textbf{Value} \\
     \midrule
     $V_{\mathrm{in}}$     & 12\,V                   & $V_{\mathrm{out}}$    & 1\,V       \\
     $T_{\mathrm{on}}$     & 100\,ns                 & $f_{sw}$               & 1.67\,MHz  \\
     $C_{\mathrm{out}}$    & 200\,$\mu$F             & $C_s$                  & 60\,$\mu$F \\
     $L$                   & 440\,nH                 & $I_{\mathrm{load}}$    & $2\!\rightarrow\!14$\,A \\
     \bottomrule
   \end{tabular*}
  }
\end{table}
\section{Conclusion and Future Work}
In this work, we demonstrate a new control approach for variable-frequency SCB converters, combining a high-speed linear controller designed in the 5S framework and a time-optimal nonlinear controller designed by PMP. The 5S model delivers precise discrete-time dynamics for high-bandwidth linear controller design, while the large-signal controller relaxes switching constraints and, with a robustness tolerance, enables seamless controller integrations. Simulations and experiments confirm that load-step recovery is over ten times faster than pure linear-controller-based designs. Future work will formally prove the time-optimal law and transient-mode logic, extend both control strategies to higher phase counts, and perform comprehensive hardware validation of the large-signal controller alongside small-signal experiments.
\section*{Acknowledgments}
The authors would like to thank Muratkhan Abdirash for his invaluable assistance in formulating the time‐optimal control theory.

\bibliographystyle{IEEEtran}
\bibliography{library}
\end{document}